\documentclass[epj]{svjour}
\usepackage{graphics}

\usepackage{graphicx}
\usepackage{amsmath}
\usepackage{color}
\usepackage{lineno}
\begin{document}
\title{Confinement and fractional charge of quarks from braid group approach to holographic principle}
\author{Janusz E. Jacak
}                     
%
%
\institute{Department of Quantum Technologies, Faculty of Fundamental Problems of Technology, Wroc{\l}aw University of Science and Technology, Wyb. Wyspia\'nskiego 27, 50-370 Wroc{\l}aw, Poland}
\date{Received: date / Revised version: date}
%
\abstract{
The so-called holographic principle, originally addressed to high energy physics, suggests more generally that the information contents of the system (measured by its entropy) scales as the event horizon surface. It has been formulated also a holographic super-string model for the anti de Sitter space, which allows for implementation of quantum gravity in the volume by only quantum boundary---hologram. The locally planar topology of the boundary of 3D space leads, however, to more reach possibilities of quantization of many particle systems according to representations of related braid groups (i.e., first homotopy groups of the configuration space for multiparticle systems). This level of freedom would be helpful in symmetry-term considerations of hypothetical locally 2D hologram properties corresponding to anyons. Specific properties of anyons on a sphere are compared with fractional charge of quarks. The confinement of quarks in hadrons is conjectured to be linked with the collective behavior of anyons precluding their individual separation.
%
} 
\maketitle
\section{Introduction}

\begin{figure}[!ht]
\centering
\scalebox{0.45}{\includegraphics{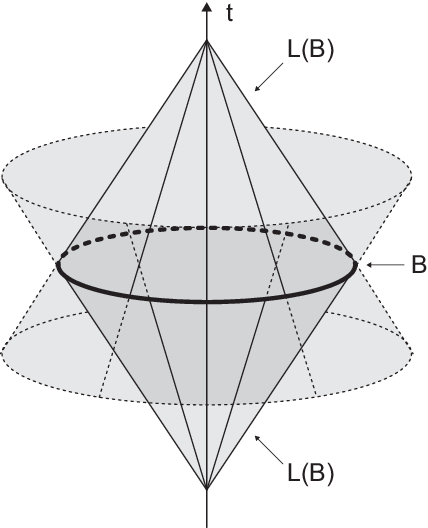}}
\caption{Two light-sheets for 2D spherical surface $B$ model ($B$ is the circle for 2D): $t$ is the time, the cones created by light rays with negative expansion define light-sheets suitable for covariant entropy bound formulated by Bousso: the entropy on each light-sheet $S(L(B))$ does not exceed the area of $B$, i.e., $A(B)$, $S(L(B)) \leq A(B)/4$ (based on Ref. \cite{Bousso2002}).}
\label{f1}
\end{figure}

The holographic principle, formulated on the basis of the Bekenstein entropy bound \cite{Bekenstein1972,tHooft2009,Bousso1999}, strongly limits locality \cite{tHooft2009,Bousso1999}.
This principle can be formulated \cite{Bousso2002} in terms of a number of independent quantum states describing the light-sheets $L(B)$ (cf. Fig. \ref{f1})---the number of states, ${\cal{N}}$, is bound by the exponential function of the area $A(B)$ of the surface $B$ corresponding to the light-sheet $L(B)$, 
\begin{equation}
{\cal{N}}[L(B)] \leq e^{A(B)/4}. 
\end{equation}
It can be also expressed equivalently \cite{Bousso2002}, that the number of degrees of freedom (or number of bits multiplied by $ ln 2$ ) involved in description of $L(B)$, cannot exceed $A(B)/4$. Even though the complete holographic theory is not constructed as of yet, this idea is considered as breakthrough on the way to quantum gravity---the holographic principle highlights the lower two-dimensional character of  the information (entropy) corresponding to any real 3D system and in that sense refers to some mysterious 2D hologram collecting all information needed to describe corresponding 3D system. 

An interesting step toward an explicit formulation of the holographic theory was done by Maldacena \cite{Maldacena1997,Witten1998}, who proposed a holographic superstring theory formulation in the anti de Sitter space. According to that model, the 3D quantum gravity could be implemented by 2D quantum nonrelativistic hologram on the edge of the anti de Sitter space (for the anti de Sitter space the boundary is a sphere at fixed time). The locally two dimensional spatial manifolds for the quantum hologram corresponding to 3D quantum relativistic systems open, however, quite new possibility for interpretation of the nature of hypothetical holographic counterparts of elementary particles. In the present letter we emphasize such a possibility referred to a specific topology of the locally 2D manifolds. It is well established in solid state physics, both experimentally \cite{Tsui1982} and theoretically \cite{Laughlin1981}, that 2D quantum dynamics results in new sorts of quantum particles besides bosons and fermions, so called anyons \cite{WilczekWS} and composite fermions \cite{jain1989}, which  elucidate fractional quantum Hall effect and Hall metal hierarchy observed in semiconductor 2D electron gas \cite{jac,Haldane1983} and, more recently, in graphene \cite{gr2,fqhe1,bil,bil2}.

The novelty of 2D quantum many particle physics follows from the exceptional reach topological structure of the multiparticle system on 2D plane. It can be expressed in algebraic topology---the homotopy group terms. The first homotopy group $\pi_1$ of the configuration space for the system of $N$ identical particles (called as the braid group) describes all possible topologically inequivalent trajectories for interchanges of particles and thus allows for recognition of their quantum  statistics. Since the braid group is especially large (infinite) for 2D space (bigger than for any other dimensions, where the braid groups are finite groups) the  new exotic types of quantum particles---anyons are allowed on the 2D plane. They are attributed to the plethora of the unitary representations of the braid group in 2D. This braid group, $\pi_1(Q_N(R^2))$ is defined  for $N$ identical particles on the plane $R^2$. $Q_N(M) = (M^N-\Delta)/S_N$ (here $M=R^2$) denotes the configuration space of indistinguishable $N$ particles on the manifold $M$, $\Delta$ is the set of diagonal points, i.e., points with coinciding particle positions (subtracted from the normal product, $M^N=M\times M\times \dots \times M$, in order to assure the particle number conservation), $S_N$ is the permutation group (the quotient by $S_N$ structure assures the quantum indistinguishability of identical particles)---for further details cf. Refs \cite{birman,jac-ws}. Though the holographic space for e.g.,  anti de Sitter space is the sphere at fixed time, not the plane $R^2$, the quantum consequences of nontrivial topology of the 2D plane maintain also for this locally two-dimensional manifold.

\section{Braid groups and their unitary representations for 2D manifolds}
\subsection{Braid groups for the plane $R^2$ and the sphere $S^2$}

For 2D plane $R^2$ the braid group was originally described by Artin \cite{Artin1947}. The Artin braid group, i.e., $\pi_1(Q_N(R^2))$ is the infinite group, with generators $\sigma_i$ (defining exchanges of neighboring particles, $i$th particle with $(i+1)$th one, at some enumeration of particles, arbitrary, however, due to particle indistinguishability), satisfying the conditions \cite{Artin1947,birman},
\begin{equation}
\sigma_i \cdot \sigma_{i+1} \cdot \sigma_i=\sigma_{i+1}\cdot\sigma_i \cdot \sigma_{i+1}, \;for\;\; 1\leq i\leq N-2,
\label{a}
\end{equation}
\begin{equation}
\label{b}
 \sigma_i \cdot \sigma_j=\sigma_j\cdot \sigma_i, \; for
\;\; 1\leq i,j \leq N-1,\;|i-2|\leq 2.
\end{equation}

For 3D sphere $S^2$ the first homotopy group $\pi_1(Q_N(S^2))$ is also complicated and infinite. Similarly as for 2D plane the braid group $\pi_1(Q_N(S^2))$ is generated by generators with properties (\ref{a}) and (\ref{b}) (the same as for $R^2$ space) but supplemented with the additional condition, 
\begin{equation}
\label{c}
\sigma_1\cdot\sigma_2 \dots \sigma_{N-2} \cdot \sigma_{N-1}^2 \cdot \sigma_{N-2} \dots \sigma_2\cdot \sigma_1=e,
\end{equation}
corresponding to a specific global property of the sphere ($e$ is a neutral element in the braid group). The latter property of generators results from the fact that a closed curve on the sphere can be interpreted in two different ways if refereed to the opposite sides of the sphere. For example, when a selected particle traverses a loop around all other particles, such a loop can be contracted via the opposite side of the sphere to a zero loop (neutral element $e$), what is, however, impossible on $R^2$---this  is illustrated in Fig. \ref{f2}.

The quantum mechanical description of a system with the configuration space $Q_N(M)$ consists in construction for a fixed time a state vector as a  multiparticle wave function transforming $Q_N(M)$ into complex numbers $C$. This function can generally be multivalued (e.g., the double-valued for fermions), which results from a change of the phase factor when traversing a loop in the space $Q_N(M)$. When making a closed loop in $Q_N(M)$, the wave function must transform according to the unitary one-dimensional representation of the first homotopy group $\pi_1(Q_N(M))$ (multidimensional representations may be linked  to multidimensional wave functions---not considered here). Therefore, in order to describe a quantization of the many-particle system, one has to choose a specific one-dimensional unitary representation of the braid group describing the phase transformations of a state vector when particles (represented by the multiparticle wave function coordinates) exchange positions according to loops defined by braids \cite{birman,jac-ws}.

The closed paths in the space $Q_N(M)$ describe exchanges of particle positions and therefore they are connected with the statistics of indistinguishable particles. The quantum statistics of particles  corresponds thus  to the irreducible unitary representation of the braid group. Note, however, that in the case of a multiply connected M manifold, even a single-particle system (thus with not defined statistics) on such multiple connected manifold $M$ may  have a complex braid group with  non-trivial unitary  representations not linked, however, to the  statistics. $S^2$ similarly as $R^2$ are simple-connected spaces, thus distinct unitary representations of the corresponding braid groups always assign various statistics and  give possible different sorts of quantum particles (with distinct statistics) allowed on these manifolds. For  $M=R^d$ ($d>2$) the corresponding braid groups are always the permutation groups \cite{Artin1947}, $\pi_1(Q_N(M)) = S_N$, with only two distinct one-dimensional unitary representations $\sigma_i\rightarrow e^{i\Theta}$, $\Theta=0,\pi$ corresponding to bosons and fermions, respectively. Nevertheless, for $R^2$ the braid group $\pi_1(Q_N(R^2))$ (the Artin braid group) is the infinite non-abelian group with infinite number of one-dimensional unitary representations $\sigma_i\rightarrow e^{i\Theta}$, $\Theta \in (-\pi,\pi]$. Each value of the parameter $\Theta$ is attributed to a different representation and enumerates different sort of quantum particles. $\Theta  = 0, \pi$  correspond to bosons and fermions, while other values of $\Theta$ describe new types of particles (possible on $R^2$)---anyons, obeying a fractional statistics \cite{WilczekWS,wu}. These new types of quantum particles in $R^2$ space turn out to be crucial in description of quasiparticle excitations in quantum Hall systems \cite{Laughlin1981,WilczekWS,Haldane1983,Hanna1989}, though are not sufficient to describe a composite fermions responsible for Hall metal and fractional quantum Hall effect, convincingly manifested themselves in the experiment. Composite fermions cannot be implemented via braid group unitary representations---these are restricted to the phase region $[0, 2\pi)$, whereas composite fermions need the phase $(2l + 1)\pi$, ($l-integer$). It is, however, possible to generalize braid groups for the plane ($R^2$) \cite{epl,jac-ws}, introducing new generators, $b_i=\sigma^l_i$, corresponding to interchange of neighboring particles but with  additional  loops  (each loop supply additional $2\pi$ phase shift which is equivalent to fictitious flux quanta pinned to composite fermions).  

\subsection{One-dimensional unitary representations of braid group for $S^2$---application to hypothetical holographic picture of quarks}

\begin{figure}[!ht]
\centering
\scalebox{0.4}{\includegraphics{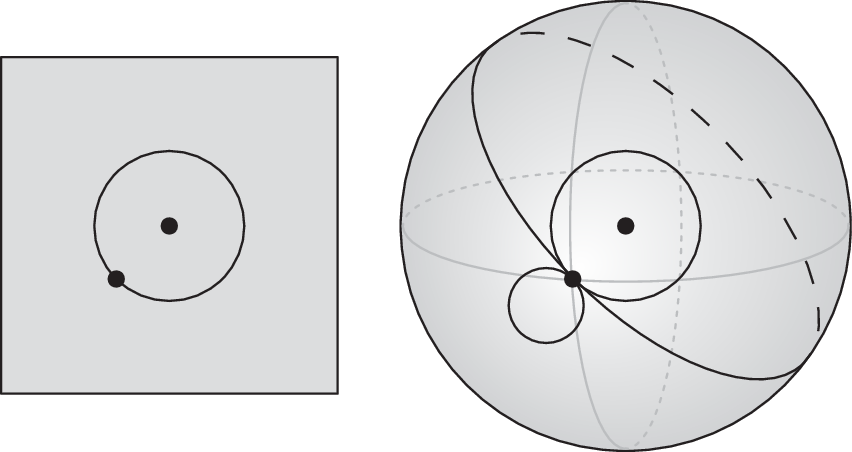}}
\caption{A loop traversed by one particle around another one cannot be contracted on $R^2$  (left) in contrary to the sphere $S^2$ where such a loop can be contracted via the opposite side of the sphere to the zero loop   (i.e., to the neutral element of the braid group) (right)}
\label{f2}
\end{figure}

For the sphere $S^2$, despite its local isomorphism with $R^2$, the one-dimensional unitary  epresentations of $\pi_1(Q_N(S^2))$ are confined to the discrete set of $e^{i\Theta}$, 
\begin{equation}
\label{d}
\Theta=\pm k\pi/(N-1), \; mod \; 2\pi, 
\end{equation}
$k = 0, 1, 2, 3, \dots , 2N - 3$ (cf. Ref. \cite{imbo1990}). This restriction imposed on unitary representations  follows from the global property of the sphere expressed by condition (\ref{c}) for generators of $\pi_1(Q_N(S^2))$. Condition (\ref{c}) plays the role in the abelization procedure, crucial in the construction of one-dimensional unitary representations \cite{imbo1990}. Thus, for the sphere $S^2$, besides bosons ($k = 0$) and fermions ($k = N - 1$), also other sorts of quantum particles are possible---anyons (for other values of $k$ suitably to $N$ according to (\ref{d})), they are, however, not so rich as for $R^2$. Moreover, for the sphere anyons may exist only for $N\geq 3$ in contrary to the plane $R^2$ with anyons  without any restriction on $N$. On the sphere two particles cannot be anyons but on the plane they can be. To be anyons on the sphere  3 particles are needed at least (it is noticeable in Fig. \ref{f2}).

Note that for $ N > 3$, $N = 3l + 1$, $l = 1, 2, \dots$,  the parameter $\Theta= \pi/3$ and $2\pi /3$ define the corresponding   anyon populations. They might be helpful   to represent holographic quarks, as it will be described below.

Let us underline that the case $S^2$ is exceptional among all locally planar closed manifolds. For the 2D torus and other closed 2D manifolds (with exception of $S^2$ including its homotopical deformations) the corresponding braid groups have only bosonic and fermionic one-dimensional unitary representations (no anyons are admitted for them) \cite{imbo1990}.

From point of view of the Aharonov-Bohm effect \cite{WilczekWS}, when two charged particles with a charge $q$ and with a magnetic field flux $\Psi$ attached to each particle  interchange   positions on the manifold $M$, the corresponding wave function acquires  the  phase shift $\Theta =q \Psi$. Thus the phase factor $\Theta$ enumerating the sort of anyons (it is the phase of the particular one-dimensional unitary representation of the braid group generator of elementary exchange $\sigma_i$, $e^{i\Theta}$) can be associated alternatively either with the charge $q$  or magnetic flux  $\Psi$ attached to the particles, provided that $q\Psi$  is conserved. In particular, for $\Theta= \pi/3$ (or $\Theta= 2\pi/3$) one can associate this value of $\Theta$ with the fractional flux $\Psi=\pi/3$ ($2\pi/3$) and the integer charge $q = 1$ (in the units $| e |$), or conversely, with the integer flux $\Psi=\pi$ (as for ordinary fermions) and fractional charge $q= 1/3$ ($2/3$)---just as  for  quarks. In this sense the fermions associated with the  fractional charge are equivalent with the anyons with the  integer charge, provided that anyons are admitted for the given manifold, like in our case the $S^2$ holographic sphere. For 3D and for higher dimensions, as well as for a  torus anyons obviously do not exist.

For the hypothetical holographic $S^2$ space the appropriate sorts of anyons (with $\Theta = \pi/3$ or $\Theta = 2\pi/3$) are, however, accessible (for $N > 3$, $N = 3l + 1, l = 1, 2, \dots$) and thus one can associate holographic quarks with these anyons assuming their integer charge but fractional flux attached to each particle or conversely---fractional charge and integer-fermionic flux, as shown above. 

The latter  would serve as the explanation of the fractional charge of quarks. Moreover, such an anyonic holographic representation would lead simultaneously to a simple explanation of the second exotic property of quarks---the confinement precluding their liberation from hadrons. The confinement of holographic quarks  follows from the fact that the number of anyons cannot be changed by one, but only by their collections with integer total flux (or equivalently total charge---just as for quarks in hadrons). This latter property of anyons can be noticed on the simplified example of anyon description within the mean-field approach upon  Chern-Simons field model \cite{Hanna1989}. The averaging of all fluxes attached to particles (anyons) results in a mean magnetic  field which for $ | e |$  charged particles (fermions) results in $n$ completely filled Landau levels for $\Theta  = \pi(1-1/n)$ sort of anyons and with $N$ particles per surface unit. All Landau levels have the same degeneracy (in the case of $R^2$), thus any change of $N$ in this picture is possible only by adding by one particle to all $n$ completely filled Landau levels created by the mean field of fluxes $\Psi$ (this mean field is dependent on $N$). Thus the smallest change of the  number $N$ of anyons is by a packet of $n$ anyons simultaneously. Note also that the similar arguments do not allow for a change of a sample area (it means the incompressibility of the quantum multiparticle state)---as such a change would modify degeneration of completely filled Landau levels, whereas the redistribution of particles is blocked by the energy gap between Landau levels.

An independent argument for specific confinement of anyons  follows also from the fractional generalization of Pauli excluding principle for anyons \cite{Haldane1991}. According to that idea, a number of allowed quantum states for eventual occupation $\Delta d$ changes with variation $\Delta N$ of number of particles as  $\Delta d = - g\Delta N$. For bosons $g = 0$, for fermions $g = 1$ and for anyons $g = \Theta /\pi$. Since $\Delta d$ and $\Delta N$ are integers, thus  $g = \Theta /\pi$ must be a rational number, and for the smallest nonzero $\Delta d = 1$ it gives the smallest $\Delta N = 1/g$ (we thus arrive with the smallest packed of n anyons for $\Theta  = d(1 - 1/n)$).

An additional interesting property of anyons described upon the Chern-Simons field model is that a residual statistical interaction beyond the averaged-mean-field is in part the 2D electric  interaction  even for free anyons \cite{Hanna1989} (this interaction does not satisfy 3D Maxwell equations, and moreover, the rest of the statistics-origin-interaction of anyons is essentially of three-particle type). The anyons manifest a Higgs-like mechanism \cite{jac} related to a specific anyonic superconductivity \cite{WilczekWS}.

The well developed mathematical formalism for consideration of anyons on the sphere $S^2$ lies in the  basis of so-called Haldane sphere model for fractional quantum Hall effect \cite{Haldane1983}. An uniform magnetic-type mean field of  Chern-Simons field piercing the sphere is the construction of a Dirac monopole in this geometry. The total flux is quantized and Landau levels are here shells of monopole harmonics \cite{Wu1976}. The Haldane sphere allows for numerical analysis of small systems with finite number of anyons including ingredients connected with their interaction which would result in choice of specific statistics due to the energy minimization. This approach has been widely used in the investigations of Hall systems \cite{jac}, but Haldane sphere geometry would even better fit to the holographic space $S^2$ suggested above. The holographic Maldacena model \cite{Maldacena1997,Witten1998} (despite being  referred to the super-string formulation and unrealistic anti de Sitter metrics) demonstrates implementation of quantum gravity by a nonrelativistic quantum hologram. This feature would be, however, of particular interest, as the anyon theory is well formulated only in the nonrelativistic case, whereas its relativistic version encountered serious problems, and they would be conveniently avoided via nonrelativistic holographic formulation.

Summarizing, we have indicated a new topology-type opportunity connected with the holographic principle and with the 2D character of the hypothetical holographic border space and have  discussed some advantages of the possible  description of holographic counterparts of quarks as anyons.

\begin{acknowledgement}
Supported by the NCN projects P.2011/02/A/ST3/00116
\end{acknowledgement}

\end{document}